\documentclass[twoside]{article}
\usepackage{epsf, amsmath, amssymb}
\input{Qcircuit}

\catcode`\@=11
\long\def\@makefntext#1{
\protect\noindent \hbox to 3.2pt {\hskip-.9pt  
$^{{\eightrm\@thefnmark}}$\hfil}#1\hfill}		

\def\thefootnote{\fnsymbol{footnote}}
\def\@makefnmark{\hbox to 0pt{$^{\@thefnmark}$\hss}}	
	
\def\ps@myheadings{\let\@mkboth\@gobbletwo
\def\@oddhead{\hbox{}
\rightmark\hfil\eightrm\thepage}   
\def\@oddfoot{}\def\@evenhead{\eightrm\thepage\hfil
\leftmark\hbox{}}\def\@evenfoot{}
\def\sectionmark##1{}\def\subsectionmark##1{}}

\oddsidemargin=\evensidemargin
\renewcommand{\thefootnote}{\fnsymbol{footnote}}

\newcounter{sectionc}\newcounter{subsectionc}\newcounter{subsubsectionc}
\renewcommand{\section}[1] {\vspace{12pt}\addtocounter{sectionc}{1} 
\setcounter{subsectionc}{0}\setcounter{subsubsectionc}{0}\noindent 
	{\tenbf\thesectionc. #1}\par\vspace{5pt}}
\renewcommand{\subsection}[1] {\vspace{12pt}\addtocounter{subsectionc}{1} 
\setcounter{subsubsectionc}{0}\noindent 
{\bf\thesectionc.\thesubsectionc. {\kern1pt \bfit #1}}\par\vspace{5pt}}
\renewcommand{\subsubsection}[1] {\vspace{12pt}\addtocounter{subsubsectionc}{1}
	\noindent{\tenrm\thesectionc.\thesubsectionc.\thesubsubsectionc.
	{\kern1pt \tenit #1}}\par\vspace{5pt}}
\newcommand{\nonumsection}[1] {\vspace{12pt}\noindent{\tenbf #1}
	\par\vspace{5pt}}

\newcounter{appendixc}
\newcounter{subappendixc}[appendixc]
\newcounter{subsubappendixc}[subappendixc]
\renewcommand{\thesubappendixc}{\Alph{appendixc}.\arabic{subappendixc}}
\renewcommand{\thesubsubappendixc}
	{\Alph{appendixc}.\arabic{subappendixc}.\arabic{subsubappendixc}}

\renewcommand{\appendix}[1] {\vspace{12pt}
        \refstepcounter{appendixc}
        \setcounter{figure}{0}
        \setcounter{table}{0}
        \setcounter{lemma}{0}
        \setcounter{theorem}{0}
        \setcounter{corollary}{0}
        \setcounter{definition}{0}
        \setcounter{equation}{0}
        \renewcommand{\thefigure}{\Alph{appendixc}.\arabic{figure}}
        \renewcommand{\thetable}{\Alph{appendixc}.\arabic{table}}
        \renewcommand{\theappendixc}{\Alph{appendixc}}
        \renewcommand{\thelemma}{\Alph{appendixc}.\arabic{lemma}}
        \renewcommand{\thetheorem}{\Alph{appendixc}.\arabic{theorem}}
        \renewcommand{\thedefinition}{\Alph{appendixc}.\arabic{definition}}
        \renewcommand{\thecorollary}{\Alph{appendixc}.\arabic{corollary}}
        \renewcommand{\theequation}{\Alph{appendixc}.\arabic{equation}}
        \noindent{\tenbf Appendix \theappendixc #1}\par\vspace{5pt}}
\newcommand{\subappendix}[1] {\vspace{12pt}
        \refstepcounter{subappendixc}
        \noindent{\bf Appendix \thesubappendixc. {\kern1pt \bfit #1}}
	\par\vspace{5pt}}
\newcommand{\subsubappendix}[1] {\vspace{12pt}
        \refstepcounter{subsubappendixc}
        \noindent{\rm Appendix \thesubsubappendixc. {\kern1pt \tenit #1}}
	\par\vspace{5pt}}

\newcommand{\textlineskip}{\baselineskip=13pt}
\newcommand{\smalllineskip}{\baselineskip=10pt}

\newcommand{\copyrightheading}[1]
	{\vspace*{-2.5cm}\smalllineskip{\flushleft
	{\footnotesize }\\
	{\footnotesize \copyright\kern }\\
	 }}

\def\abstracts#1#2#3{{
	\centering{\begin{minipage}{4.5in}\footnotesize\baselineskip=10pt
	\parindent=0pt #1\par 
	\parindent=15pt #2\par
	\parindent=15pt #3
	\end{minipage}}\par}} 

\def\keywords#1{{
	\centering{\begin{minipage}{4.5in}\footnotesize\baselineskip=10pt
	{\footnotesize\it Keywords}\/: #1
	 \end{minipage}}\par}}

\renewenvironment{thebibliography}[1]
        {\frenchspacing
	 \ninerm\baselineskip=11pt
         \begin{list}{\arabic{enumi}.}
        {\usecounter{enumi}\setlength{\parsep}{0pt}     
	 \setlength{\leftmargin 12.7pt}{\rightmargin 0pt}
         \setlength{\itemsep}{0pt} \settowidth
	{\labelwidth}{#1.}\sloppy}}{\end{list}}

\newcounter{itemlistc}
\newcounter{romanlistc}
\newcounter{alphlistc}
\newcounter{arabiclistc}

\newcommand{\fcaption}[1]{
        \refstepcounter{figure}
        \setbox\@tempboxa = \hbox{\footnotesize Fig.~\thefigure. #1}
        \ifdim \wd\@tempboxa > 5in
           {\begin{center}
        \parbox{5in}{\footnotesize\smalllineskip Fig.~\thefigure. #1}
            \end{center}}
        \else
             {\begin{center}
             {\footnotesize Fig.~\thefigure. #1}
              \end{center}}
        \fi}

\newcommand{\tcaption}[1]{
        \refstepcounter{table}
        \setbox\@tempboxa = \hbox{\footnotesize Table~\thetable. #1}
        \ifdim \wd\@tempboxa > 5in
           {\begin{center}
        \parbox{5in}{\footnotesize\smalllineskip Table~\thetable. #1}
            \end{center}}
        \else
             {\begin{center}
             {\footnotesize Table~\thetable. #1}
              \end{center}}
        \fi}

\def\pmb#1{\setbox0=\hbox{#1}
	\kern-.025em\copy0\kern-\wd0
	\kern.05em\copy0\kern-\wd0
	\kern-.025em\raise.0433em\box0}

\def\fnt#1#2{\footnotetext{\kern-.3em
	{$^{\mbox{\scriptsize #1}}$}{#2}}}

\def\fpage#1{\begingroup
\voffset=.3in
\thispagestyle{empty}\begin{table}[b]\centerline{\footnotesize #1}
	\end{table}\endgroup}

\def\runninghead#1#2{\pagestyle{myheadings}
\markboth{{\protect\footnotesize\it{\quad #1}}\hfill}
{\hfill{\protect\footnotesize\it{#2\quad}}}}
\font\tenrm=cmr10
\font\tenit=cmti10 
\font\tenbf=cmbx10
\font\bfit=cmbxti10 at 10pt
\font\ninerm=cmr9

\font\eightrm=cmr8

\def\FigName{figure}%
\newbox\captionbox
\long\def\@makecaption#1#2{%
  \ifx\FigName\@captype
    \vskip\abovecaptionskip
    \setbox\tempbox\hbox{{\figurecaptionfont #1\hskip1em #2}}
	\ifdim\wd\tempbox< 28pc
	\centerline{\box\tempbox}
	\else
	{\figurecaptionfont #1\hskip1em #2\par}
\fi\else
  	\setbox\tempbox\hbox{{\tablecaptionfont #1\hskip1em #2}}
 	\ifdim\wd\tempbox< 28pc 
	\centerline{\box\tempbox}
	\else
	{\tablecaptionfont #1\hskip1em #2\par}%
	\fi   
 \vskip\belowcaptionskip
 \fi}
\InputIfFileExists{psfig.sty}
{\typeout{^^Jpsfig.sty inputed...ok}}{\typeout{^^JWarning: psfig.sty could be be found.^^J}}
\InputIfFileExists{epsfsafe.tex}
{\typeout{^^Jepsfsafe.tex inputed...ok}}
			{\typeout{^^JWarning: epsfsafe.tex could not be found.^^J}}
\InputIfFileExists{epsfig.sty}
{\typeout{^^Jepsfig.sty inputed...ok}}{\typeout{^^JWarning: epsfig.sty could not be found.^^J}}
\InputIfFileExists{epsf.sty}
{\typeout{^^Jepsf.sty inputed...ok}}{\typeout{^^JWarning: epsf.sty could not be found.^^J}}%
\def\fps@figure{tbp}
\def\ftype@figure{1}
\def\ext@figure{lof}
\def\fnum@figure{Fig.\ \thefigure}
\def\qed{\hbox{${\vcenter{\vbox{	          
   \hrule height 0.4pt\hbox{\vrule width 0.4pt height 6pt
   \kern5pt\vrule width 0.4pt}\hrule height 0.4pt}}}$}}

\renewcommand{\thefootnote}{\fnsymbol{footnote}}  

\begin{document}
\setlength{\textheight}{8.0truein}    

\runninghead{  A Quantum Performance Simulator  $\ldots$} 
            {A. Van Rynbach, A. Muhammad, A. Mehta, J. Hussmann, and J. Kim}

\normalsize\textlineskip
\thispagestyle{empty}
\setcounter{page}{1}


\vspace*{0.88truein}

\fpage{1}
\centerline{\bf
A QUANTUM PERFORMANCE SIMULATOR BASED ON FIDELITY }
\vspace*{0.035truein}
\centerline{\bf AND FAULT-PATH COUNTING}
\vspace*{0.37truein}
\centerline{\footnotesize 
Andre Van Rynbach, Jeffrey Hussmann\footnote{Present Address: Institute for Computational Engineering and Sciences, University of Texas at Austin } , Jungsang Kim}
\vspace*{0.015truein}
\centerline{\footnotesize\it Department of Electrical and Computer Engineering, Duke University}
\centerline{\footnotesize\it Durham, NC 27705, USA}
\baselineskip=10pt
\vspace*{10pt}
\centerline{\footnotesize Abhijit C. Mehta}
\vspace*{0.015truein}
\centerline{\footnotesize\it Department of Physics, Duke University}
\centerline{\footnotesize\it Durham, NC 27705, USA}
\baselineskip=10pt
\vspace*{10pt}
\centerline{\footnotesize Ahsan Muhammad}
\vspace*{0.015truein}
\centerline{\footnotesize\it Department of Computer Science, Duke University}
\centerline{\footnotesize\it Durham, NC 27705, USA}
\vspace*{0.225truein}

\vspace*{0.21truein}
\abstracts{
Quantum performance simulators can provide practical metrics for the effectiveness of executing theoretical quantum information processing protocols on physical hardware. In this work we present a scheme to simulate the performance of fault tolerant quantum computation by automating the tracking of common fault paths for error propagation through a circuit and quantifying the fidelity of each qubit throughout the computation. Our simulation tool outputs the expected execution time, required number of qubits and the final error rate of running common fault tolerant protocols on a universal hardware, assumed to be a network of qubits with full connectivity. Our technique efficiently estimates the upper bound of error probability and provides a useful performance measure of the error threshold at low error rates where conventional Monte Carlo methods are ineffective. To verify the proposed simulator, we present simulation results comparing the execution of quantum adders which constitute a major part of Shor's algorithm. }{}{}

\vspace*{10pt}
\keywords{Quantum Performance Simulator, fidelity, quantum adder circuit, resource  tracking}
\vspace*{3pt}

\vspace*{1pt}\textlineskip	
\section{Introduction}	        
\vspace*{-0.5pt}
\noindent
Quantum computation (QC) represents a new paradigm in computing capable of solving several classically intractable problems in polynomial time, and has been an intense field of research from both a physical hardware and computer science perspective over the past two decades. Various quantum algorithms have been discovered to show the power of QC, such as Shor's factoring algorithm \cite{Shor} and Grover's unstructured search algorithm \cite{Grover}. Quantum assembly languages have been developed to formalize the process of compiling algorithms into basic executable logic elements \cite{qasm}.

Just as with classical computers, effective mapping and optimized scheduling of the logic elements on practical hardware platforms requires knowledge of the hardware architecture. QC poses additional challenges on this mapping and scheduling process arising when the physical systems representing quantum bits (qubits) suffer from decoherence processes and faulty gate operations, leading to a loss of quantum information during the computation. Fault-tolerant (FT) procedures based on quantum error-correcting codes (QECCs) were established to overcome this challenge, which comes at the cost of additional physical resources, computation time and higher levels of operational complexity. The {\em threshold theorem} proves that scalable QC is possible using error correcting codes, as long as (1) the error probability for the physical process is below a certain threshold value ({\em error threshold}), and (2) sufficient quantum and classical resources are available to implement the error correction codes necessary to ensure FT \cite{threshold}. Although detailed procedures for achieving FT are well-established, quantitative estimates of the resource overhead, computation time and resulting performance gain (in terms of reduced error rate) require an automated software tool due to the complexity in these procedures.  Previous work has proven the existence of a rigorous lower bound on the error threshold for distance 3 concatenated codes, such as the Steane code considered here \cite{Comparative, AGP}.  Our work serves as a generalization of this type of analysis by numerically finding an error threshold and tracking the individual error rates on qubits throughout the execution of a full quantum circuit.  This method overcomes the limitation of many Monte Carlo based analyses where simulating low error rates becomes prohibitively slow \cite{SteaneOverhead}.  Furthermore, our quantum performance simulator keeps accurate track of the resource and computation time overhead of running a given circuit and, in addition to tracking the error probabilities, provides a means to evaluate the effectiveness of the hardware architecture and optimize the choice of FT strategy for a finite-size quantum processor.

\section{Quantum Error Correcting Codes and Fault Tolerance}
\label{QECC_FT}
\noindent
QECCs are needed to protect the qubits from the faulty physical processes used to prepare, store, and process the qubits, and form the basis of FT quantum computation (FTQC). In general, a QECC uses $m$ physical qubits to encode $k$ logical qubits in  a way that some set of possible error events can be reliably corrected. The quantum error correction condition gives a necessary and sufficient condition for the code space $C=\{| \psi_i \rangle \}$ to correct a set of errors $E = \{E_a\}$: an error is correctible if and only if 
\begin{equation}
 \langle \psi_i | E_a^{\dagger} E_b | \psi_j \rangle = C_{ab} \delta_{ij}
 \end{equation}
for any $E_a, E_b \in E$ and any $| \psi_i \rangle, | \psi_j \rangle \in C$ for a set of real numbers $C_{ab}$~\cite{KnillLaFlamme}. Intuitively, this reflects the fact that the results of errors acting on codewords must be orthogonal in order to reliably distinguish between different possible errors, and that no information must be gained about the encoded qubit state in order to preserve quantum coherence. Typical errors on individual qubits cause arbitrary relative phase accumulation on a quantum state, but can be treated as discrete bit flip ($X$) or phase flip ($Z$) errors occurring with small probability. This is because the error states collapse into a definitive qubit state upon syndrome measurement when the errors are discretized \cite{Nielsen_Chuang}. A typical codeword for QECC is a highly entangled state satisfying certain symmetry among the constituent qubits, analogous to the parity check condition in classical linear codes.

The fact that the qubit is protected by QECC does not guarantee that one can perform FTQC. When the information is decoded from the QECC, it is no longer protected and becomes vulnerable to error. FTQC thus requires that the entire computation is performed on encoded qubits for protection against errors.  For a given physical error rate $p$, it also requires that all operations on the logical qubits keep the errors within the range correctable by the QECC to first order in $p$. Finally, the error introduced by qubit  measurement process should have probabilities of order $p^2$ \cite{Nielsen_Chuang}. We will refer to these as FT conditions.

Stabilizer codes are an important class of QECCs where the symmetry of the codewords is described by a set of Pauli operators called {\em stabilizers} \cite{Gottesman}. Valid codewords of a stabilizer code are eigenstates of every stabilizer of the code with eigenvalue of $+1$.  We focus as an example on a stabilizer code called the Steane code \cite{Steane}, where $m=7$ physical qubits are used to encode $k=1$ bit of information. This code has the capability of correcting one bit-flip error and one phase-flip error on any of the constituent qubits of an encoded state. Furthermore, it has the property that all Clifford group gates can be performed {\em transversally}, defined as applying the gate in a bit-wise fashion. Since the error from one constituent qubit does not propagate to another qubit in the same code, transversal gates automatically satisfy the FT requirement. The procedures for ensuring FT in stabilizer measurement and non-Clifford group gate necessary for universal QC is well established for the Steane code \cite{Zhou}.

Pauli operators (gates) span the operator space of a single qubit, and consist of four operators.  Pauli operators on multiple qubits are defined as tensor product of Pauli operators acting on individual qubits. A set of Pauli operators form a group called Pauli Group, denoted by $\mathcal{P}$.  Another important group is called the Clifford group $\mathcal{C}$ consists of operators (gates) that satisfy
\begin{equation}
\mathcal{C} \equiv \{U | \forall P \in \mathcal{P}, UPU^\dagger \in \mathcal{P} \}.
\end{equation}
$\mathcal{C}$ contains Pauli operators, along with Hadamard gate, CNOT, and phase gate, S.

The Gottesman-Knill theorem shows that it is possible to efficiently simulate a quantum circuit containing only Clifford group gates on a classical computer.  In order to achieve the true speed up of a universal QC, one must be able to use at least one non Clifford group gates, such as the $T=\begin{pmatrix} 1 & 0 \\ 0 & e^{i\pi / 4} \end{pmatrix}$ gate or the Toffoli gate (three qubit control-control-NOT gate, where the state of the target bit is flipped if and only if both control bits are $|1\rangle$) shown in Fig. \ref{fig:toffoli}.
\begin{figure}[tb]
\begin{equation}
\Qcircuit @C=0.65em @R=0.5em @!R {
 & \lstick{|a\rangle} &  \ctrl{1} &\qw & \rstick{|a\rangle} \\
 & \lstick{|b\rangle} & \ctrl{1} & \qw & \rstick{|b\rangle} \\
 & \lstick{|c\rangle}  & \targ & \qw &\rstick{|c \oplus ab\rangle} \\
}\nonumber 
\end{equation}
\vspace*{13pt}
\fcaption{Toffoli gate where $\oplus$ is bitwise addition. The last qubit is flipped if and only if first two qubits are both $|1\rangle$.}
\label{fig:toffoli}
\end{figure}

\subsection{Fidelity as a Performance Metric} \label{sec:fidelity}
Our performance simulator quantifies the correctness of a circuit by recording the fidelity of each qubit during computation. The fidelity between a pure state $|\psi\rangle$ and a mixed state with density matrix $\hat{\rho}$ is defined as $F(\hat{\rho},|\psi\rangle)=tr\sqrt{\langle \psi | \hat{\rho} |\psi \rangle}$ \cite{Nielsen_Chuang}.   For example, if the initial quantum state is $|\psi\rangle$, then if it is subject to a bit flip error with probability $p$, the final state density matrix $\rho$ and the corresponding fidelity are given by 
\begin{eqnarray}
\rho & = & (1-p)|\psi\rangle \langle \psi | + pX |\psi\rangle \langle \psi | X \\
F & = & \sqrt{(1-p)+p(\langle \psi | X |\psi \rangle \langle \psi | X |\psi \rangle)}.
\end{eqnarray}
Fidelity is also defined for a gate operation. If one obtained a density operator $\rho_U$ after applying a gate $U$ on an initial state $|\psi \rangle$ instead of the ideal result $U|\psi \rangle$, the fidelity of the gate operation $U$ is defined as $F(U)\equiv \min_{|\psi\rangle}F(U|\psi\rangle, \rho_U)$, where minimization is taken for all possible initial state $|\psi\rangle$. For the example of the bit flip channel above, $F=\sqrt{1-p}\simeq 1-p/2$.

\setcounter{footnote}{0}
\renewcommand{\thefootnote}{\alph{footnote}}

\section{QUIPSIM Overview}
\noindent
The main purpose of our QUantum Information Processor SIMulator (QUIPSIM) is to analyze the interaction of qubits in a quantum algorithm implementing FT procedures.  We establish performance metrics for fault tolerant quantum circuits based on the number of ancilla qubits necessary (representing resource overhead), time steps (representing execution time), and overall fidelity of the qubits at the end of computation. Fidelity is a metric chosen to represent both the quality of the qubit and the accuracy of the quantum gate operations (see Sec. 2.1).  QUIPSIM keeps track of the local error rates of qubits throughout the QC process as a way to determine error properties of logical qubits, perform resource counting, and optimize FT protocols.  The ultimate goal is to bridge the gap between theoretical proofs of FT and realistic error probabilities that must be reached for an experimental realization of a FT quantum processor using a stabilizer code.

The execution complexity of a given logic gate is dictated by the QECC chosen to achieve FT. In the Steane code, gates in the Clifford group are straightforward to perform as they can be implemented transversally. FT implementation of gates outside the Clifford group is highly complex, with increasing levels of complexity for gates with smaller rotational angles or increasing number of control bits \cite{Zhou}. Fortunately, the {\em Solovay-Kitaev theorem} proves that an arbitrary single qubit gate can be efficiently approximated to arbitrary accuracy by employing just one gate outside Clifford group, typically chosen to be the $T$ gate, to achieve universality~\cite{Nielsen_Chuang}. Our non-Clifford group gate of choice is the Toffoli gate (see Fig.~\ref{fig:toffoli}), as it is very effective in implementing arithmetic logic, such as modular exponentiation, that underlies Shor algorithm. QUIPSIM simulates a universal set of FT quantum gates made of Pauli operators, CNOT, Hadamard, and the Toffoli gate, as well as state preparation, error correction and qubit measurement. This covers all necessary operations to execute a universal FT quantum algorithm using Steane code.

Given a quantum circuit at a logical level, QUIPSIM first breaks each FT logic operation into a set of operations at the constituent level. This provides a much larger and more complex quantum circuit at one concatenation level lower than the original circuit. It performs this circuit expansion iteratively until the circuit reaches the physical level, comprised of qubit initialization, elementary logic gates, and qubit measurement operations. The error probability (or fidelity, see Sec. 2.1) and execution time of each physical operation is provided as an input to the simulator, from which the error probability and execution time of a larger block of circuit ({\em e.g.}, logic gate at a higher level of concatenation) is estimated.  Lastly, the effect of decoherence in our system is taken into account by multiplying the fidelity by a factor of $e^{- \lambda \Delta t}$ for a wait time $\Delta t$ in between gates, with a decoherence rate of $\lambda$. 

\begin{figure}[tb]
\centering
\includegraphics[width=8.2cm]{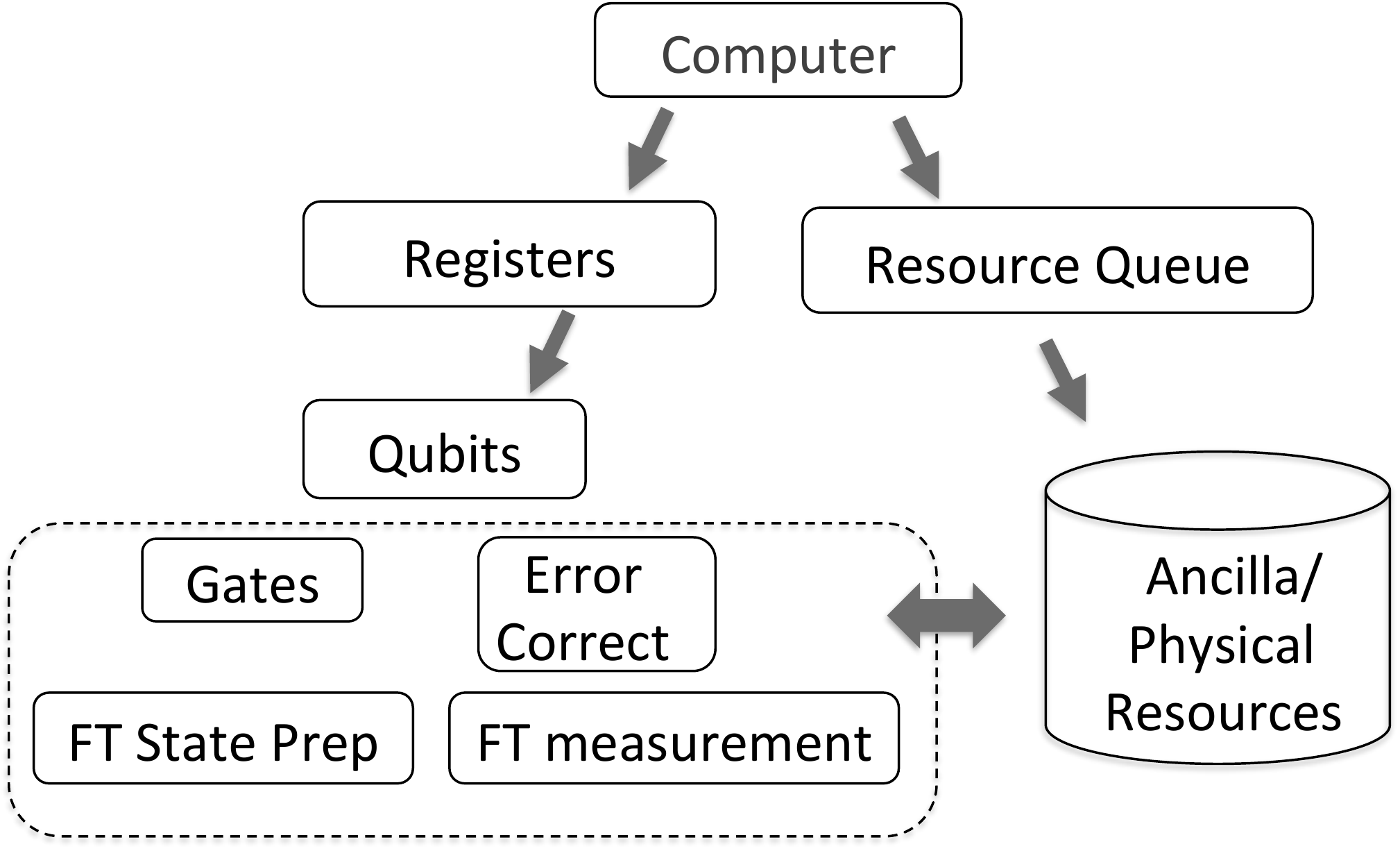}
\vspace*{13pt}
\fcaption{Code organization for the QUIPSIM program.  The computer object is made up of registers which store qubits and a globally shared resource queue of available ancilla and physical resources.  Qubit operations, including gates, error correction, state preparation, and measurement, are executed under the constraints given by the availability of ancilla and physical operational resources.  }
\label{fig:block}
\end{figure}

The code organization of  the QUIPSIM program is shown in Fig.~\ref{fig:block}.  It implements a hierarchical organization from the computer down to the individual qubits in the registers.  The overall quantum computer class contains a sub class for quantum registers to store the qubits and a resource queue to hold the available ancilla and physical resources to run parallel operations.  Under the quantum register class is the qubit class which stores the qubits inside the actual registers and performs all the logical and physical level gate operations.

\section{Implementation of Quantum Operations}
\label{Gates}
\noindent
This section summarizes the detailed procedures adopted to break a logical gate operations down into the constituent operations at a lower level of concatenation. We also describe how we account for the fidelity of the resulting qubits in each of these circuit blocks.

\subsection{Single Qubit Gates}
\noindent
The single qubit gates chosen in our approach for universal quantum computation include the Pauli operators and the Hadamard operator. Since all of these operators can be implemented transversally, each gate is replaced by the necessary bit-wise single qubit operation at the constituent level. The fidelity of each output constituent qubit is simply the product of the corresponding input constituent qubit and the fidelity of the bit-wise single qubit operation applied to the qubit.

\subsection{CNOT Gates and Error Propagation}
\begin{figure}[tb]
\centerline{\includegraphics[width=8.2cm]{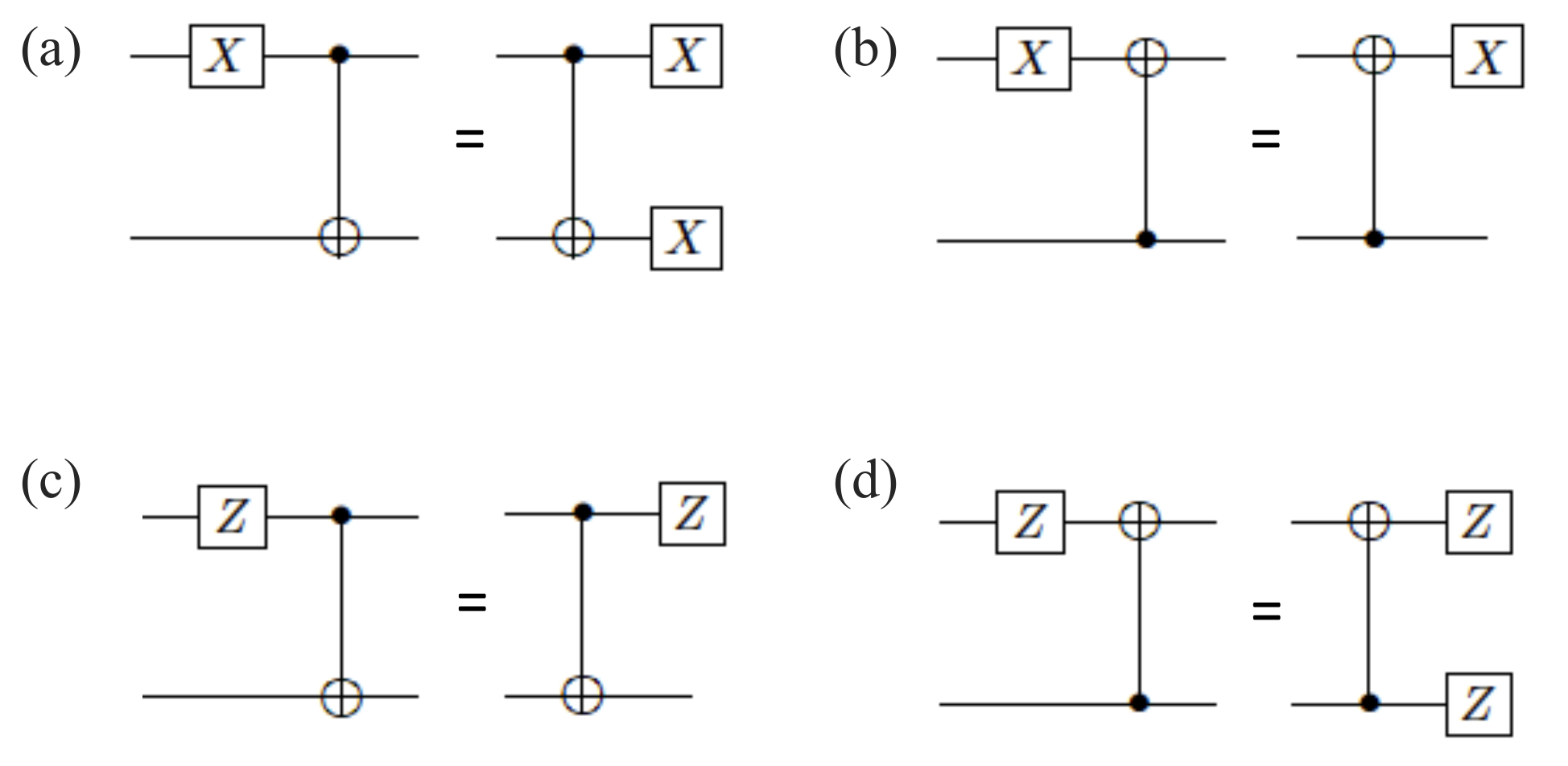}}
\vspace*{13pt}
\fcaption{Propagation of $X$ and $Z$ errors through the CNOT gates.}
\label{fig:error_prop}
\end{figure}

\noindent
CNOT gate is also transversal in Steane code, so the logical CNOT gate is replaced by bit-wise CNOT operations at the constituent level. Figure \ref{fig:error_prop} shows the circuit identities that describe the processes of "error propagation" through CNOT gate. Figure \ref{fig:error_prop}a (\ref{fig:error_prop}d) shows that $X$ ($Z$) error on the control (target) qubit propagates to the target (control) qubit, while Figure \ref{fig:error_prop}b (\ref{fig:error_prop}c) shows that $X$ ($Z$) error on the target (control) qubit is contained in the target (control) qubit and does not propagate to the control (target) qubit. In most cases, we make a simplifying assumption that the application of a CNOT gate operation degrades the fidelity of the output qubits by multiplying the fidelities of both the input qubits together and then multiplying that by the fidelity of the gate operation itself. This  provides the worst-case estimate on the fidelity of the output qubit, as some errors do not propagate through the CNOT gate to degrade the fidelity of the resulting qubits.

\subsection{Cat State Preparation}
\begin{figure}[tb]
\centerline{\includegraphics[width=8.2cm]{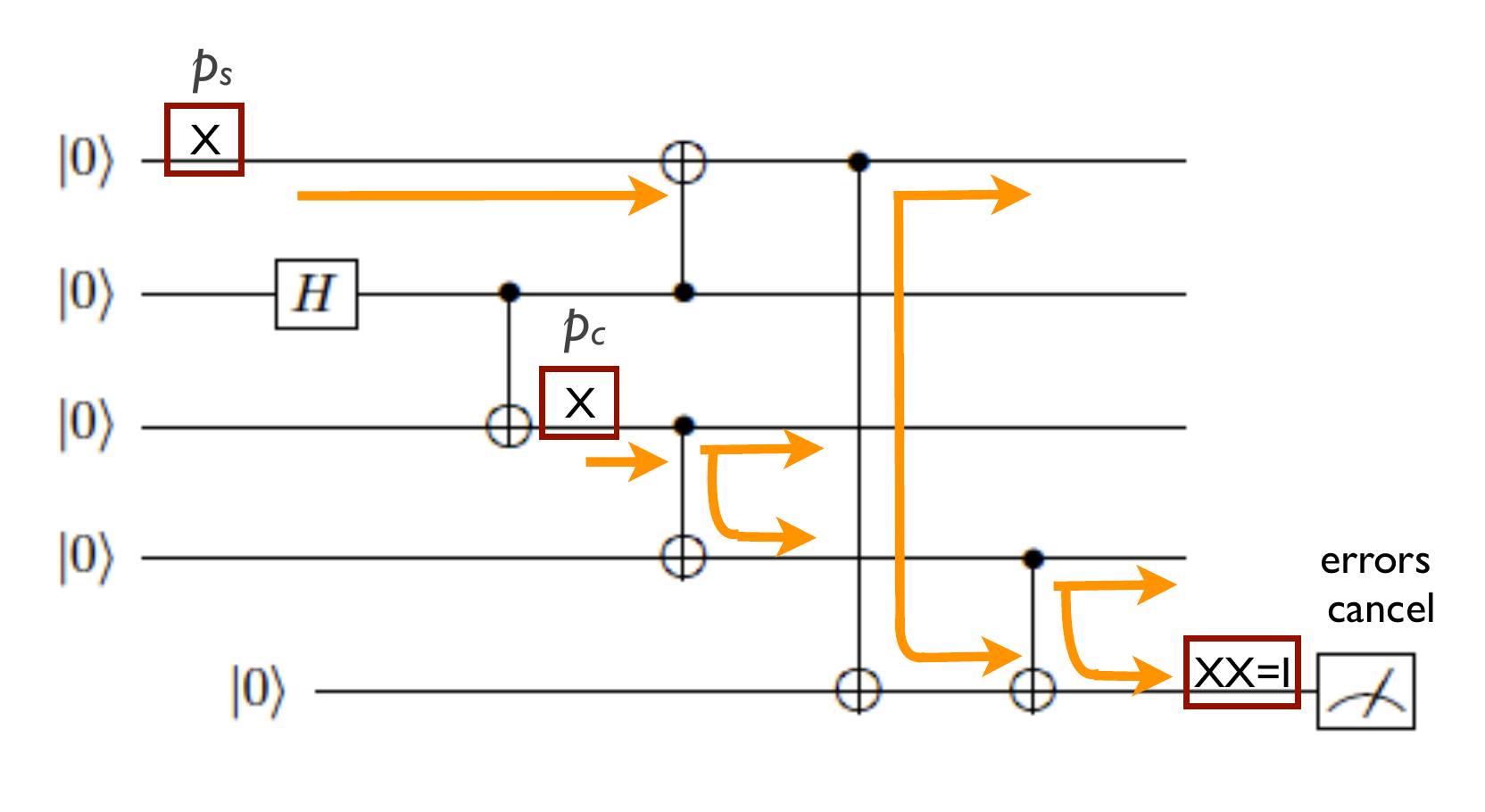}}
\vspace*{13pt}
\fcaption{Preparation of a cat state with an example of error propagation through the circuit.  The probability of two errors canceling each other, as shown here, is equal to $(l+1)p_s p_c$ for an $l$-qubit cat state.}
\label{fig:cat_state_prep}
\end{figure}

One common procedure which must be implemented in stabilizer-based FT quantum computation is preparation of maximally entangled states called cat states, used primarily for FT measurement.  The typical preparation scheme for a four qubit cat state is shown in Fig.~\ref{fig:cat_state_prep}. The first four qubits form a cat state after the Hadamard and first three CNOT gates are applied. The probability that the cat state generated contains errors is proportional to the error rate of the single gate error rate $p$, and is not good enough for use in FT measurement. A fifth verification qubit is introduced to check the parity of the first and the last qubit in the cat state, by the last two CNOT gates and the measurement. If the parity is odd, the cat state is discarded and the cat state preparation is repeated. If the parity is even, the generated cat state has an error probability of $O( p^2)$, and is ready for use.

To verify the error probability of the cat state prepared in this way, we trace the fault paths of possible  $X$ and $Z$ errors through the circuit in Fig.~\ref{fig:cat_state_prep}.  If we let $p_s$ and $p_c$ be the probability of having an error during physical state preparation and CNOT gate, respectively, we note that any single $X$ error in the process will propagate to the verification qubit and be detected. The final fidelity of an $l$-qubit cat state is derived by counting the number of ways in which the verification qubit fails to detect one or more errors remaining in the prepared cat state. Fig.~\ref{fig:error_prop} shows an example of such an occasion, where the physical state preparation error in the first qubit is cancelled at the verification qubit by an error in the first CNOT operation (modeled as $X$ error in the target qubit following the CNOT gate), resulting in two qubits flipped in the cat state. The fidelity for each of the $l$ constituent qubits in the resulting cat state is given by $f_{cat} \simeq 1-\frac{1}{2} [ (l-1)p_s ^2 + (l+1)p_s p_c + (l+1) p_c^2 ] + O(p^3) \sim 1-(3l+1)p^2/2$ when $p_s = p_c \equiv p$, satisfying the FT condition (3). We note that $Z$ errors cannot be detected in this scheme, but show in the next section that these cat states can still be used to effectively perform FT stabilizer measurements.

\subsection{State Preparation and Logical State Detection}
\label{stateprep}
\begin{figure}[tb]
\centerline{\includegraphics[width=8.2cm]{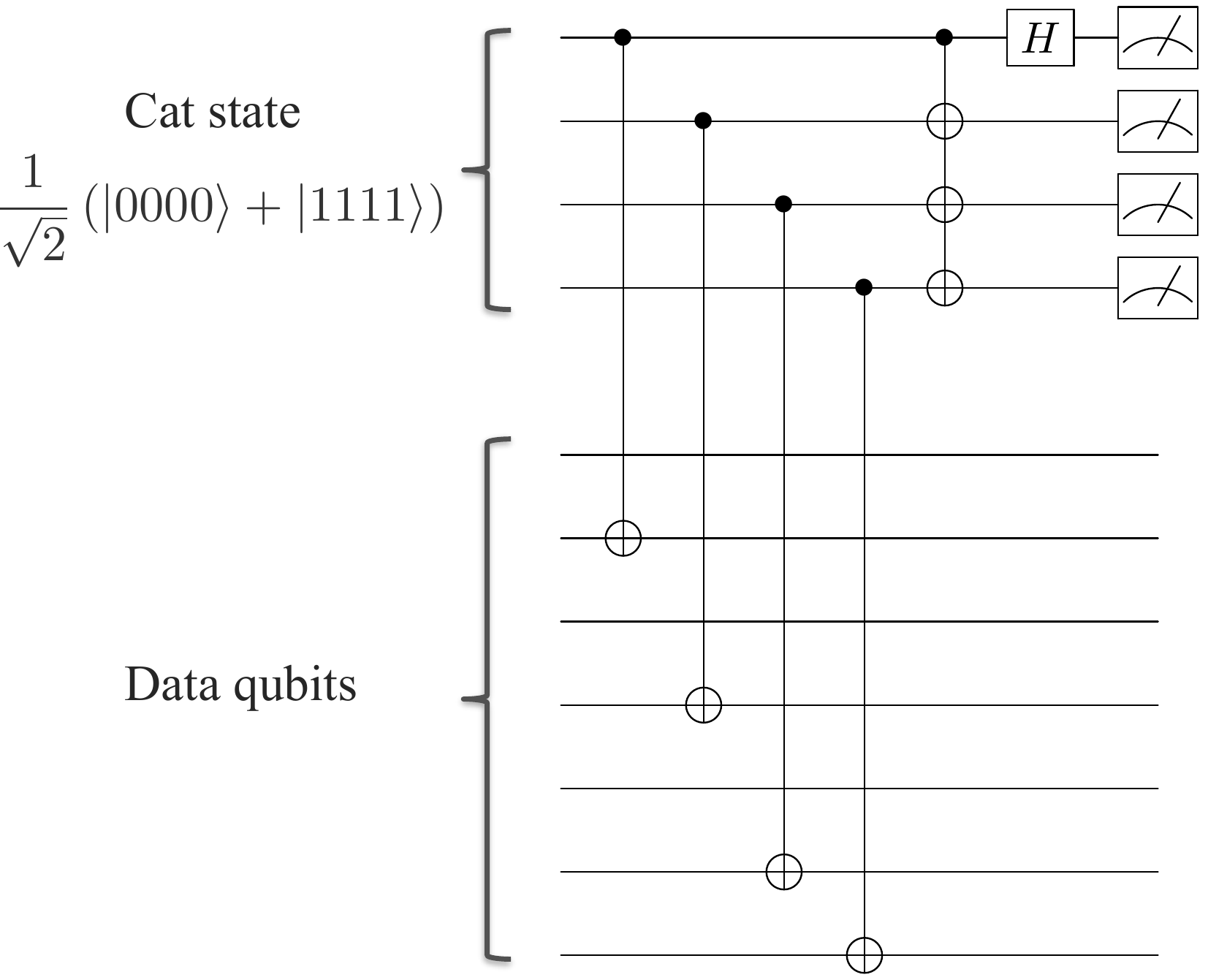}}
\vspace*{13pt}
\fcaption{FT measurement of the $g_2 = IXXIIXX$ generator of the Steane code using a  4- qubit cat state.  Stabilizer measurements such as this are used in both state preparation and in error correction.}
\label{fig:stabilizer_circuit}
\end{figure}

QUIPSIM uses the Steane encoding scheme to prepare the logical $|0\rangle$ states, achieved by measuring the six stabilizers on an arbitrary (mixed) state of seven physical qubits.  Each stabilizer measurement uses a 4-qubit cat state, as shown in Fig.~\ref{fig:stabilizer_circuit} \cite{Nielsen_Chuang}.  We choose the approach that uses six 4-qubit cat states to measure the stabilizers simultaneously rather than recycle a single cat state, to greatly reduce the execution time for the state preparation.  We note that an efficient protocol exists that uses decoding circuits to overcome the slow measurement processes that forced us to parallelize the stabilizer measurements~\cite{DiVincenzo}. After all the CNOT gates are executed, we measure the first qubit in each of the six cat states to determine the eigenvalue of the stabilizer.  If the eigenvalue -1,  we apply the logical $X$ operation to adequate constituent qubit which flips the state back to +1 eigenstate~\cite{Gottesman}.

If the cat state used for the measurement is prepared using procedure shown in Fig.~\ref{fig:cat_state_prep}, the stabilizer measurement is FT against a single $X$ error propagating into the data qubits. Any $Z$ error in either the cat state preparation or in the stabilizer measurement process does not propagate into the data qubits (due to propagation shown in Fig. \ref{fig:error_prop}c), but will result in flipping the first qubit in the cat state after the Hadamard gate leading to an error in the stabilizer readout. Error in the measurement process itself leads to a similar result in the stabilizer readout with error rate $O(p)$. Since correction operations on the data qubits are performed based on the result of the stabilizer measurement, this will lead to an error probability in the data qubit of $O(p)$, violating the FT condition. To overcome this, QUIPSIM performs the measurement of each stabilizer generator three times and uses the majority vote on the (classical) readout results to decide if the correction operation should be applied.  This reduces the error probability of  the measurement (originating either from $Z$ errors in cat states or the measurement process itself) to $p_{m} \approx 3p^2$.  The final fidelity of each physical qubit in the logical encoded state being prepared is then equal to this error probability of measurement multiplied by the error probability of the potential correction gate applied, giving a state preparation error rate of $p_{s.p.} = p(1-3p^2)$ and fidelity $f_{s.p.} = 1-p_{s.p.}/2$. This ensures that the probability of having two errors in the codeword is $O(p^2)$, satisfying the FT condition.

 After all six stabilizer measurements are performed, we must perform a logical $Z$ measurement to make sure the logical qubit is prepared as $|0\rangle$. This process is done by preparing a 3-qubit cat state and measuring the parity of the first three qubits: this parity is measured three times to reduce the measurement-induced error to $O(p^2)$.  If the $Z$ measurement indicates a result -1, then the state $|1\rangle$ is prepared: we apply the bit-wise $X$ operation on the first three constituent qubits to  flip the state back to $|0\rangle$.  The resulting fidelity of the first three qubits in a logical qubit state preparation therefore get an additional factor of $(1-p)$ from the $X$ operation. This procedure of logical $Z$ measurement also serves as the qubit state detection at the logical level.

\subsection{Error Correction} 
\label{sec:error_corr}

Part of the state preparation process, namely measuring the six stabilizers and applying associated correction factors, is a sufficient error correction procedure. QUIPSIM uses the state preparation process of Sec.~$4.4$ less the logical $Z$ measurement as the error correction procedure, which ensures that the qubit state is projected back into the code space without measuring the qubit itself. The fidelity of the qubits after error correction is calculated by adding the probability of two situations where (1) data qubits do have an error, and the syndrome measurement does not detect an error, and (2) data qubits had no error but the syndrome detection detects one, and a correction procedure was applied. This estimation accurately reduces the remaining error probability after error correction is applied to $O(p^2)$. All other situations lead to  error probabilities of $ \sim O(p^3)$, and can be safely ignored in the limit of low physical error probabilities.  QUIPSIM implements this by calculating the probability that the stabilizer measurements during error correction have an error.  Since measurements are repeated three times, the final fidelity is reduced to $f\approx 1-9p_{meas}^2 $.  While this is just an approximate expression, the implementation by the simulator computes a more accurate value by finding the measurement failure rate $p_{meas}$ for each of the three measurements, and combining them to get the probability that the error correction mistakenly finds an error when none is present or does not detect an error when one is present.

\subsection{Fault tolerant Toffoli gate}

\begin{figure}[tb]
\centering
\includegraphics[width=8.2cm]{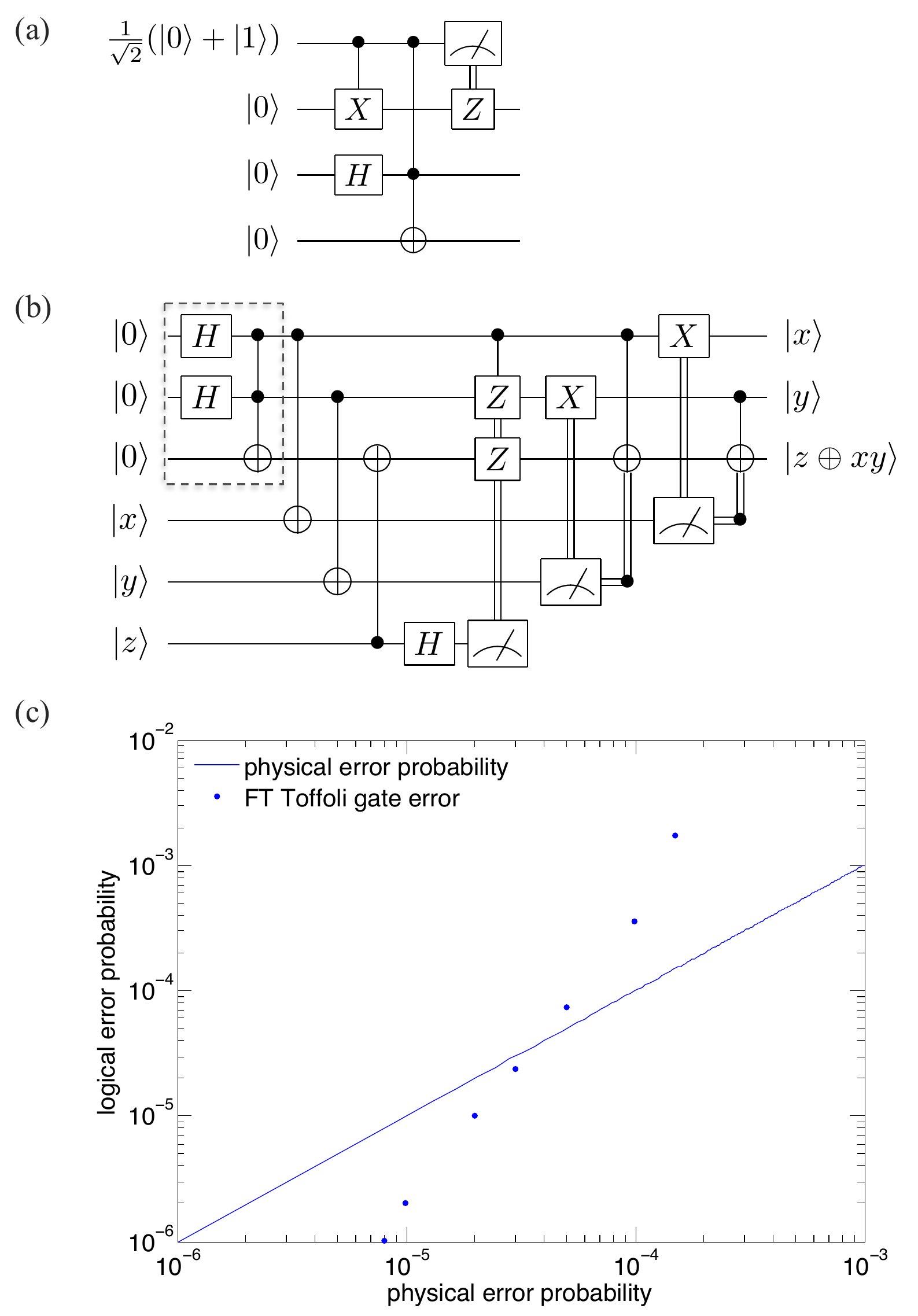}
\vspace*{13pt}
\fcaption{A FT Toffoli gate. The Toffoli and Hadamard gates in the box in circuit in (b) can be replaced with the operations shown in (a). (c) Error probability of the FT Toffoli gate as a function of the physical error probability.}
\label{fig:full_toffoli}
\end{figure}

Universal quantum computation requires at least one gate outside the Clifford group; here we choose the Toffoli gate.  Unlike the Clifford group, a fault tolerant version of the Toffoli gate cannot be implemented transversally.  A fault tolerant version of the Toffoli requires a 7-qubit cat state and three ancilla qubits prepared in a known state $|000\rangle$, as shown in Fig.~\ref{fig:full_toffoli}a \cite{Zhou}. The top qubit is the 7-qubit cat state, and the CNOT, Hadamard, $Z$ and measurement are performed using the FT procedures described above. The Toffoli gate shown here is actually a bit-wise Toffoli performed on each constituent qubits. Bit-wise Toffoli itself is not a FT implementation of Toffoli gate, but in this circuit we are  measuring the eigenvalue of CNOT operation performed between the third and the fourth qubit, by performing bit-wise controlled-CNOT with the 7-qubit cat state. The bit-wise Toffoli can be further broken down into lower-level circuit if it is still encoded. We have to assume that our physical hardware can perform a Toffoli gate at the physical level, where experimental demonstrations exist~\cite{Monz}. The resulting three-qubit state replaces the first two Hadamard gates and the Toffoli gate in the circuit shown in Fig. \ref{fig:full_toffoli}b, which shows a FT implementation of the Toffoli gate. All remaining operations consist of CNOTs, controlled-$Z$ gates (which can be implemented with one CNOT and two Hadamard gates), qubit measurements and single qubit gates, which can be simulated fault tolerantly by QUIPSIM.

Figure Fig. \ref{fig:full_toffoli}c shows the error probability of the FT Toffoli gate circuit with one level of encoding as a function of the physical error probability. It shows that below the physical error probabilities of $p \sim 4\times10^{-5}$, the encoded gate shows much better error probabilities than the uncoded one.  This was taken in the no memory error limit where the decoherence rate is much longer than the gate time.  Since the FT Toffoli gate has the highest error rate of all the gates in the universal set described here, this error value can be thought of as a threshold error rate since the FT implementation of any gate decreases the overall error rate when the physical rate is lower than this value.

\section{Quantum adder circuits}
\label{sec:sim}
The most celebrated application of quantum computation is Shor's algorithm, which involves modular exponentiation (ME) and quantum Fourier transform (QFT) \cite{Shor}.  An $n$-bit ME is a classical arithmetic circuit, but requires $O(n^3)$ operations compared to the $O(n^2)$ operations for QFT and dominates the computation time. The algorithmic optimization of ME circuit has been considered, and the performance is dictated by the ability to implement efficient quantum adders \cite{vanMeter}. Several authors  have proposed optimized versions of various known adder circuits to fit into quantum domain~\cite{Cuccaro,Draper,Draper2,Gossett}, reducing the number of resource-intensive Toffoli gates and depth of the circuit. As an example application of QUIPSIM, we compare the execution time, resource requirements and error properties of two adders: quantum ripple carry adder (QRCA) and the quantum carry look-ahead adder (QCLA). Optimized QRCA features $2n+O(1)$ Toffoli gates and has depth $2n+O(1)$ for $n$-bit addition~\cite{Cuccaro}. QCLA  features  $5n-O(\log n)$ Toffoli gates and has $O(2\log n)$ depth~\cite{Draper}. 

\section{Performance Evaluation}

In this section we show some performance simulation examples demonstrating the capabilities of QUIPSIM.  Although QUIPSIM is capable of independently setting the operation times and error probabilities for each physical gate type, we made the simplifying assumption to normalize every physical gate time to 1 unit with an associated error probability of $p$ in the examples shown here to determine the more generic behavior of the circuits. We also simplify the memory error by setting the decoherence rate to $\lambda =0$ in order to isolate the effects of gate error in a system with no memory errors. Physical measurement times are assumed to be 200 times the gate time, to reflect the realities of many experimental situations where state readout is significantly slower than qubit gates.  One important assumption we made was regarding the connectivity of the qubits in the system.  In any experimental setup, detailed physical implementation of a two qubit gate between a pair of qubits in the system necessarily depends on their relative location, providing additional constraints in the performance measures.  One should should make assumptions about how to realize two qubit gates between distant pairs of qubits.  In our model, we make the simplifying assumption that a two qubit gate can applied between any pair of qubits in the system, i.e. full connectivity.

\begin{figure}[tb]
\centerline{\includegraphics[width=8.2cm]{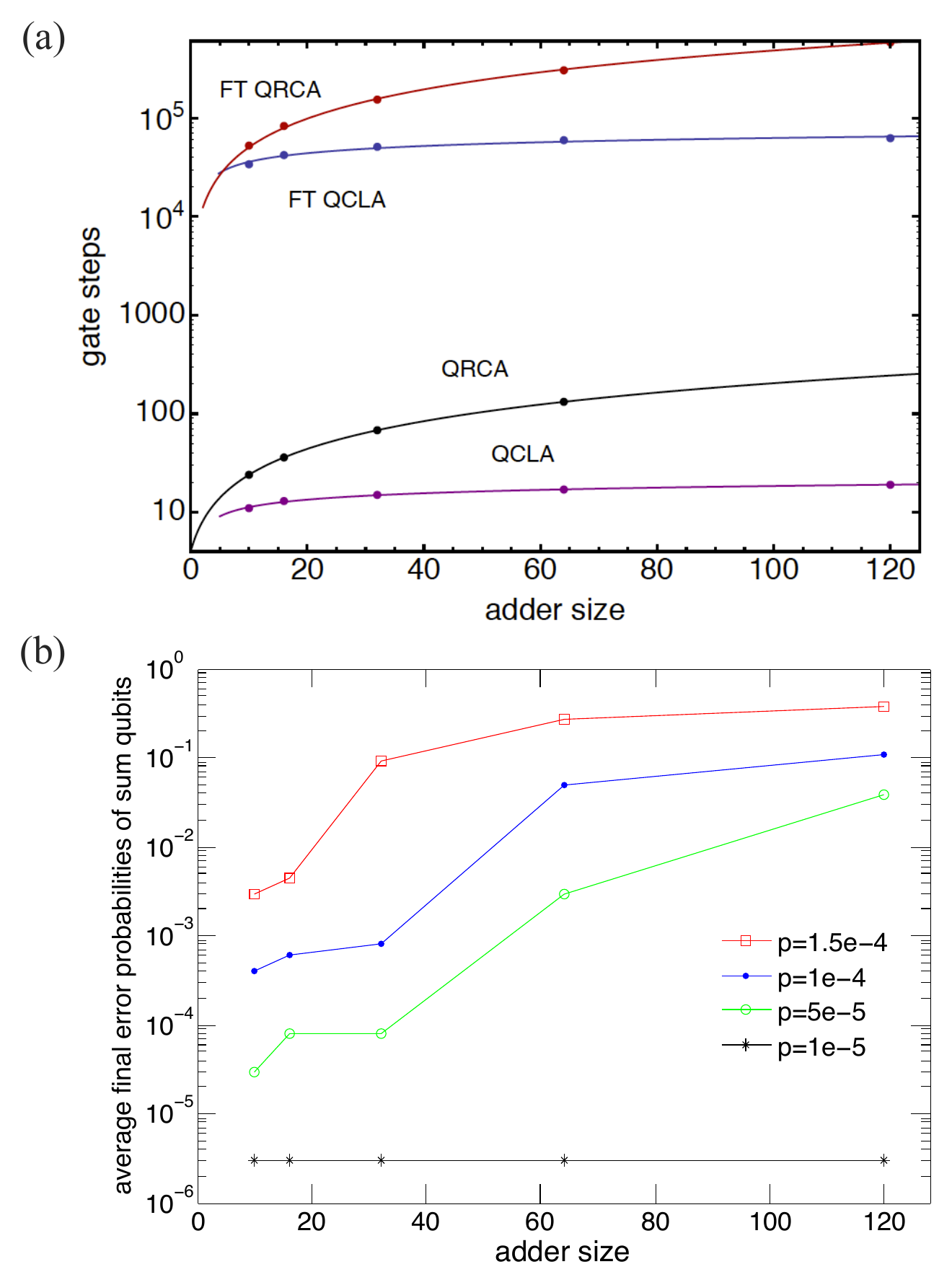}}
\fcaption{Simulation of FT QCLA and ripple carry adder circuit.  (a) Execution time of the QRCA and QCLA circuits with and without FT encoding, in units of physical gate times. (b) Average error probabilities of the output qubits for the FT QCLA circuit.  The final error probability is independent of adder size when the gate error rate is $p=1\cdot 10^{-5}$ since this is effectively below threshold.}
\label{fig:Carry_Adders}
\end{figure}

The performance of QRCA and QCLA circuits is evaluated with and without the use of FT encoding.  Fig.~\ref{fig:Carry_Adders}a shows the execution time of the two circuits measured in units of physical gate times. While the fault tolerant encoding of the circuits gives a substantial increase in the execution time (around three orders of magnitude), it does not change the nature of the circuit depth and hence the QCLA circuit still scales logarithmically in the adder size with approximately the same scale factor as the unencoded adder.  Fig.~\ref{fig:Carry_Adders}b shows the average error probability of the output qubits for the FT QCLA circuit as a function of the adder size $n$. At error probabilities below threshold ($p_{th} \sim 4 \times 10^{-5}$), the final error rate of the circuit is much lower than the physical error rate, as FT behavior would suggest. The transition above threshold is seen in cases where sharp decline in final fidelities is seen with increasing adder size.

\section{Conclusion}
QUIPSIM is a software tool capable of simulating the performance of a complete set of FTQC protocols in the presence of faulty gates and low error rates.  QUIPSIM tracks the fidelity of each qubit over the course of the circuit using two basic approaches: (1) in state preparation and error correction blocks, it explicitly computes the number of fault paths that lead to erroneous outcome to estimate the fidelity, and (2) in Clifford group gates, it performs the error estimation by keeping track of the worst-case scenario of how errors propagate through these gates.  This approach provides a conservative estimate of the error behavior of the overall circuit, but is very effective in low-error systems where other approaches like Monte-Carlo simulations take a very long time to converge. We used QUIPSIM to test the performance of two types of quantum adder circuits of various sizes, employing Steane code for the FT procedure.

QUIPSIM has the flexibility to add more realistic hardware models by defining the execution time and error probabilities for each gate type. We can further improve the FT protocol by utilizing  more efficient stabilizer measurement processes. QUIPSIM can be used to study the relationship between physical gate parameters and the FT execution of a quantum circuit to optimize FT procedures. One limitation of the current version is that the qubits are assumed to be fully connected. An important extension of this work is to evaluate the performance of quantum circuits on physical hardware with added constraints like the connectivity of qubits and availability of resources, leading to the ability to optimize the hardware architecture of quantum processors based on experimental constraints.

\nonumsection{Acknowledgements}
\noindent
We would like to thank Byung-Soo Choi and Rachel Noek for helpful discussions. This work was supported by the NSF and IARPA

\nonumsection{References}

\end{document}